# The relation between black hole spin and star formation in massive star-forming galaxies

Yongyun Chen(陈永云),[1]★ Qiusheng Gu(顾秋生),[2]★ Junhui Fan(樊军辉),[3] Dingrong Xiong(熊定荣) [,4] Xiaoling Yu(俞效龄),[1] Nan Ding(丁楠) [5] and Xiaotong Guo(郭晓通) [6]

[1]*College of Physics and Electronic Engineering, Qujing Normal University, Qujing 655011, P. R. China*
[2]*School of Astronomy and Space Science, Nanjing University, Nanjing 210093, P. R. China*
[3]*Center for Astrophysics,Guang zhou University, Guang zhou 510006, China*
[4]*Yunnan Observatories, Chinese Academy of Sciences, Kunming 650011, China*
[5]*School of Physical Science and Technology, Kunming University, Kunming 650214, P. R. China*
[6]*School of Mathematics and Physics, Anqing Normal University, Anqing 246011, P. R. China*

## ABSTRACT

It has always been believed that feedback from active galactic nuclei (AGNs) has an important impact on star formation in massive galaxies. Black hole spin is an important physical parameter of AGN. We use a large sample of massive star-forming galaxies to study the effects of AGN on star formation. Our main results are as follows: (i) there are significant correlations between black hole spin and star formation rate, specific star formation rate, and star formation activity parameter for massive star-forming early- and late-type galaxies, respectively. These results indicate that the spin of supermassive black holes regulates the star formation of massive star-forming early- and late-type galaxies. (2) The slopes of the relationship between black hole spin and star formation rate, specific star formation rate, and star formation activity parameter for massive star-forming early- and late-type galaxies are similar within the error range. These results imply that the mechanism of black hole spin regulating star formation may be similar in massive star-forming early-type and late-type galaxies.

## 1 INTRODUCTION

Galaxies in the local universe are roughly divided into two categories: blue star-forming galaxies and red quiescent galaxies. According to their visual morphologies, galaxies are widely classified as elliptical (E) galaxies, lenticulars galaxies (S0), spirals galaxies, and irregular (Irr) galaxies (Hubble 1926). It is widely believed that the star-forming activity in galaxies is closely related to their morphological type. Gas-poor elliptical galaxies and lenticular galaxies have low star formation rates (SFR lower than $\sim 1\,M_\odot\,yr^{-1}$), while gas-rich spirals and Irr galaxies have high SFRs ($\sim 20\,M_\odot\,yr^{-1}$, Kennicutt 1983; Gao & Solomon 2004; Calvi et al. 2018; Nersesian et al. 2019). E galaxies are considered to be amongst the most massive, old, and red systems (Bernardi et al. 2003; González Delgado et al. 2015; Nersesian et al. 2019). Conversely, the spiral galaxies are mostly bluer, young, and ongoing star formation activity systems.

Many studies have found that the SFR of galaxies exhibits a bimodal distribution (e.g. Wetzel, Tinker & Conroy 2012; Trussler et al. 2020; Kalinova et al. 2021; Sampaio et al. 2022). By combining two fundamental properties of galaxies, SFR and stellar mass ($M_*$), we can obtain information about their current rate of gas conversion into stars. In the case of star-forming galaxies, these two parameters were found to be closely related, occupying a different region in the SFR–$M_*$ diagram, commonly referred to as the main sequence of star formation (SFMS) of the galaxies (e.g. Elbaz et al. 2007; Noeske et al. 2007; Wuyts et al. 2011; Whitaker et al. 2012) or 'the blue cloud'. On the contrary, quiescent galaxies show a weak relationship between SFR and $M_*$, occupying the region below SFMS, forming a 'red sequence'. The area between the blue cloud and red sequence is defined as a 'green valley' (Wyder et al. 2007). Blue clouds are mainly composed of late-type galaxies (spirals and irregular galaxies), while red sequences are mainly composed of early-type galaxies (E and S0). This bimodal phenomenon has been deeply studied in previous studies (e.g. Strateva et al. 2001; Blanton et al. 2003; Baldry et al. 2004; Taylor et al. 2015). However, several studies have reported the presence of early-type galaxies with ongoing star formation, as well as late-type galaxies that have ceased star formation activity (e.g. Rowlands et al. 2012; Vulcani et al. 2015; Bitsakis et al. 2019; Cano-Díaz et al. 2019).

Through observation and numerical simulation, it is found that most of the massive galaxies are early-type galaxies (e.g. Cappellari et al. 2013). In addition, many studies have found that the mass of the central supermassive black hole is closely related to the stellar mass of the bulge (e.g. Ferrarese & Merritt 2000; Häring & Rix 2004; Kormendy & Ho 2013). Therefore, it is expected that each massive galaxy has a supermassive black hole at its galactic centre. In recent years, many studies have found that various feedback processes are crucial for the formation of galaxies (e.g. Silk & Rees 1998; Croton et al. 2006; Kang, Jing & Silk 2006; Keller, Wadsley & Couchman 2016). The galaxies generated by the simulation without feedback are too massive to match the observations (Oser et al. 2010). Due to the existence of supermassive black holes, active galactic nucleus (AGN) feedback plays a dominant role at the high halo mass end (Croton et al. 2006; Kang et al. 2006), which is so strong in massive galaxies that it can quench the entire galaxy (Springel, Di Matteo & Hernquist 2005; Khalatyan et al. 2008). According to the accretion rate, the feedback of AGN mainly includes two modes: quasar and jet modes (Fabian 2012; Heckman & Best 2014). The different observations present both the negative (e.g. Cano-Díaz et al. 2012, 2019; Lammers et al. 2023) and positive feedback evidence (e.g. Elbaz et al. 2009; Gaibler et al. 2012; Zinn et al. 2013). Martín-Navarro et al. (2018) suggested that the supermassive black hole regulates star formation in massive galaxies. Almost all massive galaxies, including our own, have central black holes with masses ranging from millions to billions of solar masses. The growth of these black holes releases a lot of energy, powering quasars and other weaker AGN. A small portion of this energy, if absorbed by the host galaxies, can affect the star formation of the host galaxies. At present, it is unclear whether the spin of a massive black hole affects the star formation of the host galaxies.

★ E-mail: ynkmcyy@yeah.net (YC); qsgu@nju.edu.cn (QG)

In this article, we investigate the relationship between black hole spin and SFR, as well as the specific SFR and star formation activity parameter for massive star-forming galaxies. The second part displays the samples; The third part is the results and discussion; The fourth part is the conclusions.

## 2 THE SAMPLE

### 2.1 The sample of massive star-forming galaxies

We choose a large sample of broad-line AGN with reliable redshift, black hole mass, bolometric luminosity, and morphological classification of host galaxies. First, we consider the sample of Zhuang & Ho (2023). Zhuang & Ho (2023) analysed 14 574 type 1 AGNs with $z \leq 0.35$ selected by Liu et al. (2019) from the seventh data release of the Sloan Digital Sky Survey (SDSS DR7). They obtained the black hole mass and bolometric luminosity of AGN, and morphological classification of AGN host galaxies. The black hole mass is estimated by the virial method using the full width at halfmaximum of broad H$\beta$ and the AGN continuum luminosity at 5100 Å ($L_{5100}$). The bolometric luminosity ($L_{bol}$) of AGN is estimated by using the $L_{5100}$, $L_{bol} = 9.8 L_{5100}$ (McLure & Dunlop 2004). Zhuang & Ho (2023) used the GALFITM to obtain the structure parameter of AGN host galaxies, Sérsic index ($n$). They defined late-type galaxies as $n < 2$, while early-type galaxies as $n > 2$. Secondly, we consider that these AGNs have reliable stellar mass and SFRs. We crossmatch these AGNs with the catalogue of Chang et al. (2015) to obtain stellar mass and global SFR. Chang et al. (2015) obtained stellar mass and global SFR by using MAGPHYS to fit the photometric spectral energy distributions (SED) of both SDSS and *Wide-field Infrared Survey Explorer*. Meanwhile, in order to calculate the spin of the black hole, we consider the source with 144 MHz radio flux from the ongoing Low-Frequency ARray Two-metre Sky Survey Data Release 2 (Shimwell et al. 2022). We obtain 151 MHz radio flux from 144 MHz radio flux using $f_\nu \propto \nu^{-\alpha}$ ($\alpha = 0.8$, Shimwell et al. 2022). Finally, we consider that the host galaxies of these AGNs are star-forming galaxies. The star-forming galaxies are defined as $\log SFR > 0.86 \times \log M_* - 9.19$ (Chang et al. 2015). Meanwhile, we find that these star forming galaxies have high stellar masses ($9.5 < \log M_*/M_\odot < 11.5$). Thus, we obtain a total of 714 massive star-forming galaxies (408 early-type galaxies and 306 late-type galaxies). The sample is shown in Table 1 and Fig. 1.

### 2.2 The spin of black hole

The spin of a black hole can be calculated using the following formula (Daly 2019)

$$\frac{f(j)}{f_{max}} = \left(\frac{L_j}{g_j L_{Edd}}\right) \left(\frac{L_{bol}}{g_{bol} L_{Edd}}\right)^{-A}$$
$$j = \frac{2\sqrt{f(j)/f_{max}}}{f(j)/f_{max} + 1}, \quad (1)$$

where $L_j$ is the beam power, $L_{Edd}$ is the Eddington luminosity ($L_{Edd} = 1.3 \times 10^{38}(M_{BH}/M_\odot)$), $g_j = 0.1$, $g_{bol} = 1$, and $A = 0.43$ are adopted (Daly 2019). The black hole spin obtained by this method can be compared with the black hole spin obtained by other methods, such as the X-ray reflection method (Reynolds 2014). The spin of GX 339−4 is $0.94 \pm 0.02$ using the X-ray reflection method (Miller et al. 2009), and the spin is $0.92 \pm 0.06$ using the method of Daly (2019). Thus, the spin obtained by the method of Daly (2019) is consistent with that obtained by the X-ray reflection method. Daly (2019) used the above equation-(1) to estimate the black hole spin of 753 AGNs (includes radio-quiet and radio-loud AGNs) and 4 stellar-mass galactic black holes. Daly (2019) suggested that there is good agreement between the spin values obtained by using the above methods and those obtained with the X-ray reflection method for AGN. Due to the limitations of observation, it is very difficult to directly measure the spin of large samples using the X-ray reflection method. Therefore, we use the method of Daly (2019) to measure the spin of our sample.

### 2.3 The beam power

The beam power can be estimated by the following formula (Cavagnolo et al. 2010),

$$L_j \approx 5.8 \times 10^{43} \left(\frac{L_{radio}}{10^{40} \text{erg s}^{-1}}\right)^{0.70} \text{erg s}^{-1} \quad (2)$$

where $L_{radio}$ is the radio luminosity at 200 MHz in units of erg s$^{-1}$. The 200 MHz radio luminosity is estimated by using the formula $L_\nu = 4\pi d_L^2 S_\nu$ and $d_L(z) = \frac{c}{H_0}(1+z) \int_0^z [\Omega_\Lambda + \Omega_m(1+z')^3]^{-1/2} dz'$, where $d_L$ is the luminosity distance (Venters, Pavlidou & Reyes 2009). We extrapolate 150–200 MHz flux using $S_\nu \propto \nu^{-\alpha}$ and make a $K$-correction for the 200 MHz radio flux using $S_\nu = S_\nu^{obs}(1+z)^{\alpha-1}$ and $\alpha = 0.8$ (Cassaro et al. 1999). Many authors used the above equation to estimate the beam power of radio-quiet AGN and radio-loud AGNs (e.g. Cavagnolo et al. 2010; Cheung et al. 2016; Mezcua, Suh & Civano 2019; Chen et al. 2020; Singha et al. 2023; Igo et al. 2024). We also calculate the beam power of our sample using equation (2). A Lambda-cold dark matter cosmology with $H_0 = 70$ km s$^{-1}$ Mpc$^{-1}$, $\Omega_\Lambda = 0.73$, and $\Omega_m = 0.27$ is adopted.

## 3 RESULT AND DISCUSSION

### 3.1 The distribution of physical parameter

It is generally believed that star formation activity in galaxies is closely related to their morphological types. Early-type galaxies are

**Table 1.** The sample of massive star-forming galaxies.

| Name (1) | RA (2) | Dec. (3) | redshift (4) | log $M_*$ (5) | log SFR (6) | log sSFR (7) | log $L_{bol}$ (8) | log $M_{BH}$ (9) | $S_{151}$ (10) | log$L_j$ (11) | $f_j/f_{max}$ (12) | $j$ (13) | Morph (14) |
|---|---|---|---|---|---|---|---|---|---|---|---|---|---|
| J073631.83 + 383058.4 | 114.13261 | 38.516217 | 0.073 | 10.67 | 0.542 | −10.128 | 44.52 | 6.64 | 1.299 | 41.97 | 0.021 | 0.281 | Late |
| J073632.76 + 312215.4 | 114.13652 | 31.370945 | 0.122 | 10.97 | 0.612 | −10.358 | 44.56 | 7.37 | 1.923 | 42.41 | 0.021 | 0.283 | Late |
| J073646.66 + 393255.4 | 114.19441 | 39.548713 | 0.107 | 10.81 | 0.167 | −10.643 | 44.62 | 8.46 | 2.707 | 42.43 | 0.005 | 0.14 | Early |
| J073956.02 + 402816.3 | 114.98341 | 40.471183 | 0.061 | 10.15 | 0.397 | −9.753 | 44.15 | 6.62 | 3.446 | 42.15 | 0.046 | 0.41 | Late |
| J073956.26 + 280144.1 | 114.98441 | 28.028918 | 0.081 | 10.15 | 0.587 | −9.563 | 44.39 | 7.84 | 2.84 | 42.27 | 0.01 | 0.195 | Early |
| J074306.07 + 402040.4 | 115.77528 | 40.344549 | 0.178 | 10.94 | 1.572 | −9.368 | 43.76 | 6.79 | 3.691 | 42.85 | 0.277 | 0.824 | Early |
| J074324.40 + 255157.0 | 115.85165 | 25.865837 | 0.126 | 10.46 | 0.477 | −9.983 | 43.48 | 6.77 | 2.526 | 42.51 | 0.172 | 0.707 | Early |
| J074646.21 + 402302.2 | 116.69256 | 40.383942 | 0.073 | 10.56 | 0.332 | −10.228 | 44.4 | 7.8 | 0.927 | 41.86 | 0.004 | 0.125 | Early |
| J074940.92 + 375508.3 | 117.42049 | 37.918982 | 0.117 | 11.17 | 0.747 | −10.423 | 45 | 7.06 | 1.267 | 42.25 | 0.014 | 0.236 | Late |
| J074942.06 + 303951.3 | 117.42527 | 30.664249 | 0.157 | 11.21 | 0.622 | −10.588 | 44.96 | 8.43 | 1.093 | 42.4 | 0.003 | 0.117 | Early |
| J074948.33 + 264734.2 | 117.45138 | 26.792828 | 0.132 | 10.78 | 1.127 | −9.653 | 44.66 | 7.16 | 12.719 | 43.03 | 0.106 | 0.588 | Early |
| J075017.49 + 270304.1 | 117.57289 | 27.051145 | 0.141 | 11.2 | 1.267 | −9.933 | 44.24 | 8.07 | 2.91 | 42.63 | 0.019 | 0.272 | Early |
| J075037.94 + 304908.6 | 117.65808 | 30.819045 | 0.158 | 11.36 | 1.272 | −10.088 | 45.38 | 8.9 | 3.564 | 42.77 | 0.003 | 0.107 | Early |
| J075356.05 + 265323.9 | 118.48354 | 26.88996 | 0.134 | 11.07 | 0.712 | −10.358 | 43.86 | 7.6 | 2.261 | 42.52 | 0.04 | 0.385 | Late |
| J075456.75 + 355627.7 | 118.73645 | 35.941037 | 0.077 | 10.77 | 0.697 | −10.073 | 44.57 | 7.66 | 3.354 | 42.28 | 0.011 | 0.204 | Early |
| J075701.32 + 372419.1 | 119.25551 | 37.405292 | 0.12 | 11.06 | 1.007 | −10.053 | 45.24 | 8.6 | 0.967 | 42.19 | 0.001 | 0.072 | Early |
| J075751.20 + 345921.8 | 119.46333 | 34.989384 | 0.07 | 10.71 | 0.997 | −9.713 | 45.04 | 8.24 | 10.079 | 42.56 | 0.006 | 0.153 | Early |
| J075956.61 + 421222.9 | 119.98586 | 42.206362 | 0.132 | 10.53 | 0.867 | −9.663 | 43.98 | 6.64 | 1.743 | 42.43 | 0.102 | 0.581 | Late |
| J080118.49 + 282921.6 | 120.32703 | 28.489333 | 0.078 | 10.76 | 0.437 | −10.323 | 43.21 | 6.28 | 4.758 | 42.4 | 0.327 | 0.862 | Early |
| J080303.63 + 515729.1 | 120.76512 | 51.958082 | 0.07 | 9.87 | 0.477 | −9.393 | 43.75 | 6.12 | 1.703 | 42.02 | 0.098 | 0.571 | Late |
| J080344.31 + 323042.6 | 120.93463 | 32.511826 | 0.209 | 10.66 | 1.752 | −8.908 | 45.11 | 6.92 | 1.861 | 42.76 | 0.049 | 0.422 | Late |
| J080352.99 + 263123.4 | 120.97079 | 26.523155 | 0.046 | 10.2 | 0.292 | −9.908 | 44.29 | 6.6 | 9.763 | 42.29 | 0.057 | 0.452 | Late |
| J080407.40 + 391927.5 | 121.03084 | 39.324303 | 0.163 | 10.92 | 1.737 | −9.183 | 45.48 | 8 | 14.38 | 43.21 | 0.024 | 0.3 | Early |
| J080459.80 + 503731.2 | 121.24917 | 50.625324 | 0.174 | 11.48 | 1.127 | −10.353 | 45.58 | 8.48 | 1.313 | 42.52 | 0.002 | 0.096 | Early |
| J080508.50 + 524017.9 | 121.28542 | 52.671631 | 0.157 | 10.12 | 1.517 | −8.603 | 44.54 | 8.04 | 2.244 | 42.62 | 0.014 | 0.237 | Early |
| J080537.37 + 362522.1 | 121.40571 | 36.422794 | 0.089 | 10.69 | 0.592 | −10.098 | 44.68 | 7.26 | 1.853 | 42.19 | 0.013 | 0.226 | Early |
| J080740.99 + 390015.3 | 121.92081 | 39.004246 | 0.023 | 10.33 | 1.107 | −9.223 | 43.16 | 6.2 | 44.449 | 42.32 | 0.321 | 0.857 | Early |
| J080752.27 + 383211.0 | 121.9678 | 38.536388 | 0.067 | 10.88 | 1.062 | −9.818 | 44.76 | 7.73 | 4.784 | 42.3 | 0.008 | 0.181 | Early |
| J080807.13 + 563832.4 | 122.0297 | 56.642338 | 0.099 | 10.2 | 0.877 | −9.323 | 43.94 | 6.51 | 0.811 | 42.01 | 0.048 | 0.42 | Early |
| J080820.76 + 361148.7 | 122.0865 | 36.196853 | 0.083 | 10.27 | 0.277 | −9.993 | 44.67 | 7.63 | 1.049 | 41.98 | 0.005 | 0.14 | Late |
| J080833.38 + 540521.2 | 122.1391 | 54.089229 | 0.285 | 11.06 | 1.292 | −9.768 | 44.18 | 7.76 | 0.511 | 42.57 | 0.027 | 0.319 | Early |
| J081013.02 + 345136.9 | 122.55423 | 34.860239 | 0.083 | 10.67 | 1.482 | −9.188 | 44.88 | 8.21 | 3.22 | 42.32 | 0.004 | 0.128 | Early |
| J081121.40 + 405451.8 | 122.83917 | 40.914381 | 0.067 | 10.3 | 1.502 | −8.798 | 44.46 | 7.6 | 5.316 | 42.34 | 0.014 | 0.237 | Early |
| J081218.84 + 362620.5 | 123.0785 | 36.439029 | 0.122 | 10.65 | 0.872 | −9.778 | 44.95 | 8.06 | 1.583 | 42.35 | 0.005 | 0.141 | Late |

*Notes.* Columns (1) is the name of the sources; column (2) is the right ascension in decimal degrees; column (3) is (delineation) in decimal degrees; columns (4) is the redshift; columns (5) is the stellar mass; columns (6) is SFR; columns (7) is specific SFR; columns (8) is the bolometric luminosity; columns (9) is the black hole mass; columns (10) is the 151 MHz flux in units mJy; columns (11) is the beam power in units erg s$^{-1}$; columns (12) is the is the spin function; columns (13) is the back hole spin; and columns (14) is the morphology of galaxies: Late is the late-type galaxies; and Early is the early-type galaxies. This table is available in its entirety in machine-readable form. The full table is available as Supporting Information.

considered to be amongst the most massive, old, and mostly inactive systems (e.g. Bernardi et al. 2003; González Delgado et al. 2015; Nersesian et al. 2019). Conversely, late-type galaxies are mostly bluer, actively star-forming systems.

The distribution of physical parameters is shown in Fig. 2. The average SFRs of early- and late-type galaxies are log SFR$_{Early}$ = 0.94 ± 0.52 and log SFR$_{Late}$ = 0.98 ± 0.48, respectively. Late-type galaxies tend to have a higher average SFR than early-type galaxies. Through a non-parametric Kolmogorov–Smirnov (K-S) test, there is no significant difference in SFRs between early- and late-type galaxies ($P = 0.38$, significant probability $P < 0.05$).

The average specific SFRs of early- and late-type galaxies are log sSFR$_{Early}$ = −9.88 ± 0.52 and log sSFR$_{Late}$ = −9.75 ± 0.45, respectively. Late-type galaxies tend to have a higher average specific SFR than early-type galaxies. Through a nonparametric K-S test, there is a significant difference in specific SFRs between early- and late-type galaxies ($P = 0.0002$). These results suggest that late-type galaxies have stronger star formation activity than early-type galaxies.

The average black hole masses of early- and late-type galaxies are log M$_{BH,Early}$ = 7.71 ± 0.57 and log M$_{BH,Late}$ = 7.12 ± 0.59, respectively. Early-type galaxies tend to have a higher average black hole mass than late-type galaxies. Through a non-parametric K-S test, there is a significant difference in black hole mass between early- and late-type galaxies ($P = 4.26 \times 10^{-28}$).

The average black hole spins of early- and late-type galaxies are log j$_{Early}$ = −1.66 ± 0.26 and log j$_{Late}$ = −1.47 ± 0.24, respectively. Late-type galaxies tend to have higher average black hole spins than early-type galaxies. Through a non-parametric K-S test, there is a significant difference in black hole spin between early- and late-type galaxies ($P = 8.03 \times 10^{-14}$). From the above results, we find that the early-type galaxies with high black hole mass tend to have low black hole spin. King, Pringle & Hofmann (2008) also discovered such a tendency. This result is consistent with the assumption that the high-mass supermassive black hole is formed by more isotropic chaotic accretion or the merging of smaller black holes (e.g. Volonteri et al. 2005; Sesana et al. 2014; Fiacconi, Sijacki & Pringle 2018).

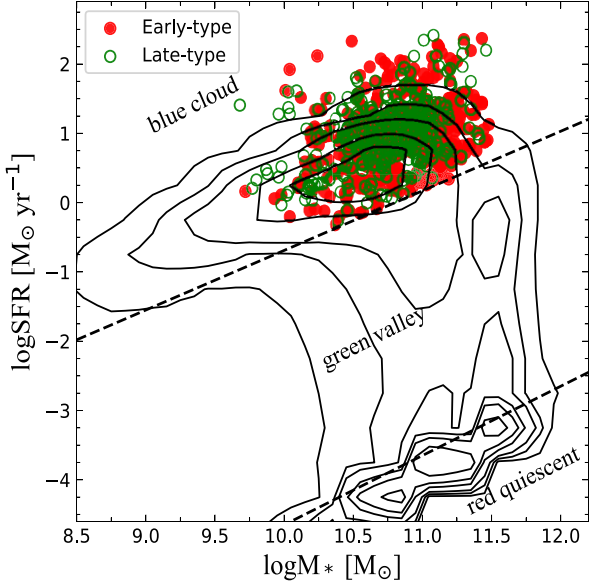

**Figure 1.** The sample of massive star-forming galaxies. The red dot is early-type galaxies, and the green dot is late-type galaxies. The contour is the sample of Chang et al. (2015). The two dashed lines divide galaxies into blue galaxies (star-forming) and red galaxies, with the green valley galaxies located between them.

### 3.2 Relation between star formation rate and specific star formation rate and black hole spin

The co-evolution of supermassive black holes and host galaxies has always been a hot issue in the formation and evolution of galaxies. Theory and observations show that the feedback of AGN plays an important role in the coevolution of supermassive black holes and host galaxies (e.g. Cano-Díaz et al. 2012; Bischetti et al. 2022; Chen et al. 2022). The feedback of AGN mainly includes radiative and jet modes (Heckman & Best 2014). According to the theoretical model of jet formation, the spin of the black hole enhances the relativistic jets (Blandford & Znajek 1977). Some observations also show that the black hole activity depends on the spin of the black hole (e.g. Narayan & McClintock 2012; Steiner, McClintock & Narayan 2013; Ünal & Loeb 2020). Therefore, the black hole spin can be used as an indicator of black hole activity.

Fig. 3 shows the relation between the spin of black holes and SFR (left panel) and specific SFR (right panel) for massive star-forming galaxies. The red dot is early-type galaxies, and the green dot is late-type galaxies. We use least-square linear regression to analyse the relationship between the spin of black holes and SFR for massive star-forming early- and late-type galaxies, respectively. In the entire paper, if the $p$-value of the null hypothesis (i.e. there is no correlation between two quantities) is $p \leq 0.05$, we consider the correlation to be significant. There is a significant correlation between the spin of the black hole and SFR for massive star-forming early-type galaxies ($r = 0.29$ and $p = 1.95 \times 10^{-9}$),

$$\log \text{SFR} = (0.57 \pm 0.09) \log j + (1.89 \pm 0.15). \quad (3)$$

There is also a significant correlation between the spin of black holes and SFR for massive star-forming late-type galaxies ($r = 0.27$ and $p = 1.04 \times^{-6}$),

$$\log \text{SFR} = (0.56 \pm 0.11) \log j + (1.79 \pm 0.16). \quad (4)$$

From equations (3) and (4), we find that the slope of the relationship between SFR and black hole spin in early- and late-type galaxies is consistent within the error range. We find that there is still a significant correlation between black hole spin and SFR when using partial correlation analysis for early-type ($r_{xy,M_*} = 0.30$ and $p = 1.0 \times 10^{-9}$) and late-type ($r_{xy,M_*} = 0.33$ and $p = 3.8 \times 10^{-9}$) galaxies, excluding the influence of stellar mass. At the same time, we also use the least-square regression to analyse the relationship between the spin of black holes and specific SFR (sSFR = SFR/$M_*$) for massive star-forming galaxies. There is a significant correlation between specific SFR and spin of black holes for massive star-forming early-type galaxies ($r = 0.27$ and $p = 1.81 \times 10^{-8}$),

$$\log \text{sSFR} = (0.53 \pm 0.09) \log j + (-9.01 \pm 0.16). \quad (5)$$

There is also a significant correlation between the spin of the black hole and specific SFR for massive star-forming late-type galaxies ($r = 0.33$, $p = 2.97 \times 10^{-9}$),

$$\log \text{sSFR} = (0.63 \pm 0.10) \log j + (-8.81 \pm 0.15). \quad (6)$$

From equations (5) and (6), we also find that the slope of the relationship between specific SFR and black hole spin in early- and late-type galaxies is consistent within the error range. There is still a significant correlation between black hole spin and SFR when using partial correlation analysis for early-type ($r_{xy,M_*} = 0.29$ and $p = 1.91 \times 10^{-9}$) and late-type ($r_{xy,M_*} = 0.32$ and $p = 8.49 \times 10^{-9}$) galaxies, excluding the influence of stellar mass.

### 3.3 Relation between star formation activity parameters and black hole spin

The amplitude evolution of $M_*$–SFR can be recalculated based on the time-scale of star formation activity. Models usually predict that galaxies have $M_*/\text{SFR} \sim t_{\text{Hubble}}$, where $t_{\text{Hubble}}$ is *Hubble* time (Davé 2008). In order to better quantify the star formation activity of galaxies, we introduce star formation activity parameters as $\alpha_{\text{sf}} \equiv (M_*/\text{SFR})/(t_{\text{Hubble}}(z) - 1\text{Gyr})$ (Davé 2008; Chen et al. 2016), where $t_{\text{Hubble}}(z)$ is the *Hubble* time at the redshift of the galaxies, and 1 Gyr is subtracted to explain the fact that star formation mainly occurred after reionization, the low $\alpha_{\text{sf}}$, i.e. rapid star formation. In physics, this can be considered as the fraction of the *Hubble* time (minus Gyr) that a galaxy needs to have formed stars at its current rate in order to produce its current stellar mass. The distributions of star formation activity parameters of early- and late-type galaxies are shown in the left panel of Fig. 4. The average star formation activity parameters of early- and late-type galaxies are $\log \alpha_{\text{sf,Early}} = -0.16 \pm 0.50$ and $\log \alpha_{\text{sf,Late}} = -0.29 \pm 0.44$, respectively. Early-type galaxies tend to have higher average star formation activity parameters than late-type galaxies. These results further indicate that early-type galaxies have weak star formation activity than late-type galaxies. Through a non-parametric K-S test, there is a significant difference in star formation activity parameters between early- and late-type galaxies ($P = 0.0001$).

The relation between the spin of black holes and star formation activity parameter for massive star-forming galaxies is shown in the right panel of Fig. 4. We find a significant correlation between the spin of the black hole and star formation activity parameter for massive star-forming early-type galaxies ($r = -0.27$ and $p = 4.57 \times 10^{-8}$),

$$\log \alpha_{\text{sf}} = (-0.51 \pm 0.09) \log j + (-0.99 \pm 0.15). \quad (7)$$

There is also a significant correlation between the spin of the black hole and star formation activity parameter for massive star-forming late-type galaxies ($r = -0.33$, $p = 3.02 \times 10^{-9}$),

$$\log \alpha_{\text{sf}} = (-0.62 \pm 0.10) \log j + (-1.20 \pm 0.15). \quad (8)$$

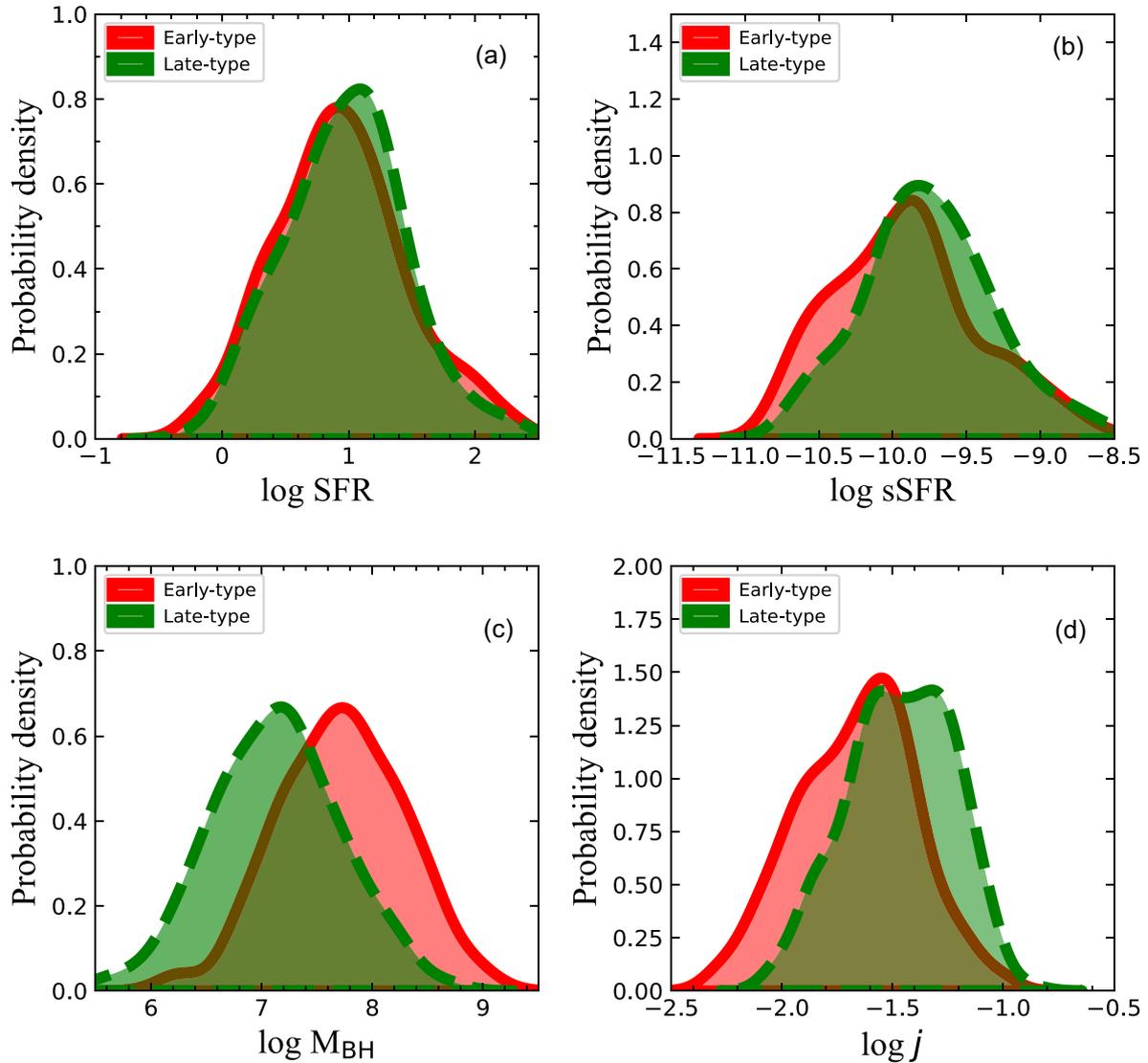

**Figure 2.** The distributions of physical parameters for massive star-forming galaxies. The red solid line is the massive star-forming early-type galaxies. The green dashed line is the massive star-forming late-type galaxies. (a) The distributions of SFR. (b) The distributions of specific SFRs. (c) The distributions of black hole mass. (d) The distributions of black hole spin.

From the results of Figs 3 and 4, we can find that the spin of supermassive black holes regulates star formation in massive star-forming early- and late-type galaxies. If the spin of the supermassive black hole enhances the relativistic jet, these results further suggest that the relativistic jet may enhance star formation in massive star-forming early- and late-type galaxies. Elbaz et al. (2009) discovered that jets promote star formation through the detailed study of the quasar HE0450−2958. Zinn et al. (2013) found that AGN with pronounced radio jets exhibit a much higher SFR, which implies that positive AGN feedback plays an important role (jet-induced star formation). Gaibler et al. (2012) found jet-induced star formation through numerical simulation of galaxy evolution.

From equation (7) and (8), we also find that the slope of the relationship between star formation activity parameter and black hole spin in early- and late-type galaxies is consistent within the error range. In summary, the slopes of the relationship between black hole spin and SFR, specific SFR, and star formation activity parameters in early- and late-type galaxies are consistent within the error range. Zhuang & Ho (2023) found that early- and late-type galaxies follow a similar $M_{BH} - M_*$ relation. Our results suggest that the mechanism by which black hole spin regulates star formation activity may be similar in early- and late-type galaxies.

Although jets can locally trigger star formation (positive feedback, e.g. Schutte & Reines 2022), simulations generally show that the long-term net effect of jets is to quench star formation (negative feedback) by disrupting the cooling flows that supply fuel for star formation (e.g. Bourne & Sijacki 2017; Su et al. 2021; Huško & Lacey 2023; Talbot, Sijacki & Bourne 2024). Star formation regulation due to AGN feedback can also occur through quasar winds, which exhibit a dependence on black hole spin in terms of both power and angular distribution (e.g. Ishibashi 2020). Similar to the behaviour of jets, AGN winds have been observed to locally enhance star formation while globally suppressing it (e.g. Mercedes-Feliz et al. 2023), with stronger star formation suppression associated with higher spin values (Bollati et al. 2024).

It is possible that the observed correlations arise from processes that simultaneously promote both high star formation and high black hole spin, without a direct causal link between them. For instance, large reservoirs of cold gas can fuel both star formation and the black hole activity, leading to a correlation between star formation and AGN luminosity (e.g. Rosario et al. 2013; Ward et al. 2022). If the accretion process is coherent, it can generate high black hole spins, which would then appear correlated with star formation.

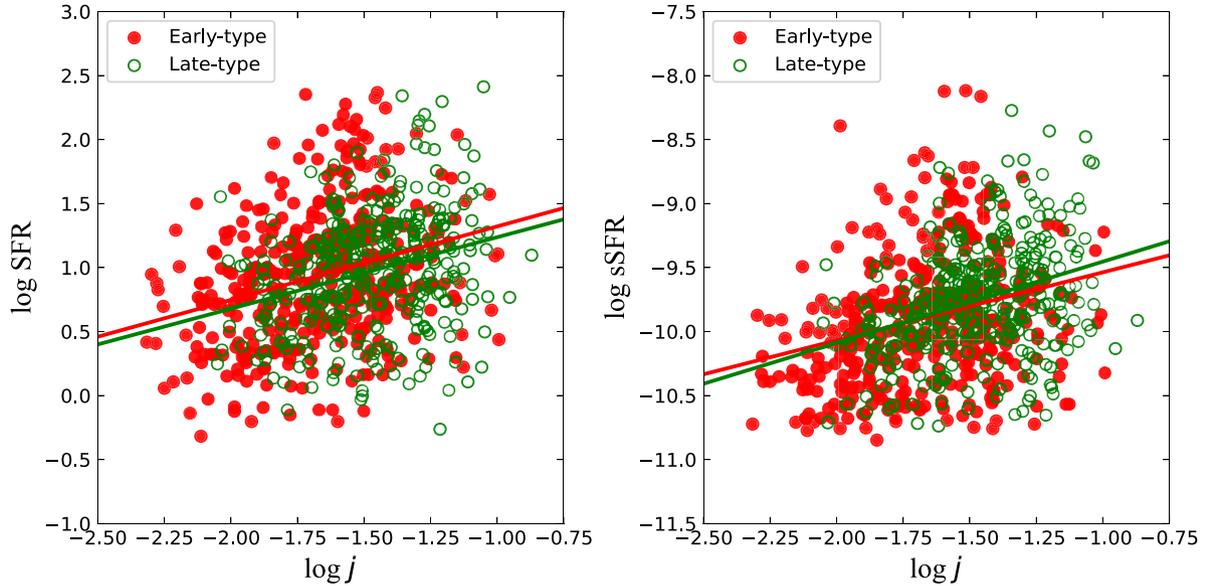

**Figure 3.** Correlation between spin of black hole and SFR (left panel) and specific SFR (right panel) (sSFR = SFR/$M_*$) for massive star-forming galaxies. The red solid line is the best-fitting line for massive star-forming early-type galaxies. The green solid line is the best-fitting line for massive star-forming late-type galaxies. The red dot is a massive star-forming early-type galaxies. The green dot is a massive star-forming late-type galaxies.

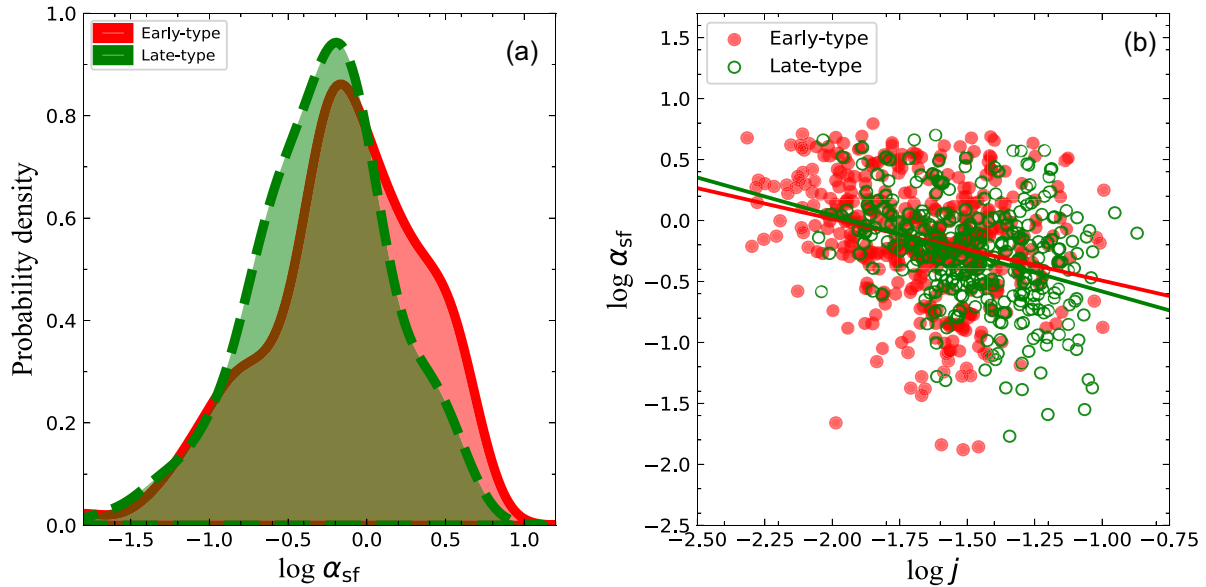

**Figure 4.** (a) The distribution of star formation activity parameter for massive star-forming galaxies. (b) The relation between star formation activity parameter and black hole spin for massive star-forming galaxies. The red solid line is the best-fitting line for massive star-forming early-type galaxies. The green solid line is the best-fitting line for massive star-forming late-type galaxies. The red dot is a massive star-forming early-type galaxies. The green dot is a massive star-forming late-type galaxies.

### 3.4 Relation between star formation rate and Eddington ratio

The distribution of Eddington ratio ($f_{\rm Edd} = L_{\rm bol}/L_{\rm Edd}$) of early- and late-type galaxies is shown in the left panel of Fig. 5.

The average Eddington ratios for early- and late-type galaxies are $\log L_{\rm bol}/L_{\rm Edd}|_{\rm Early} = -1.34 \pm 0.54$ and $\log L_{\rm bol}/L_{\rm Edd}|_{\rm Late} = -0.76 \pm 0.52$, respectively. Early-type galaxies exhibit lower average Eddington ratios compared to late-type galaxies. Additionally, late-type galaxies demonstrate stronger star formation activity relative to early-type galaxies. Our findings may suggest that accretion plays a significant role in enhancing star formation. A nonparametric K-S test reveals a statistically significant difference in Eddington ratios between early- and late-type galaxies, with a $p$-value of $P = 2.74 \times 10^{-24}$.

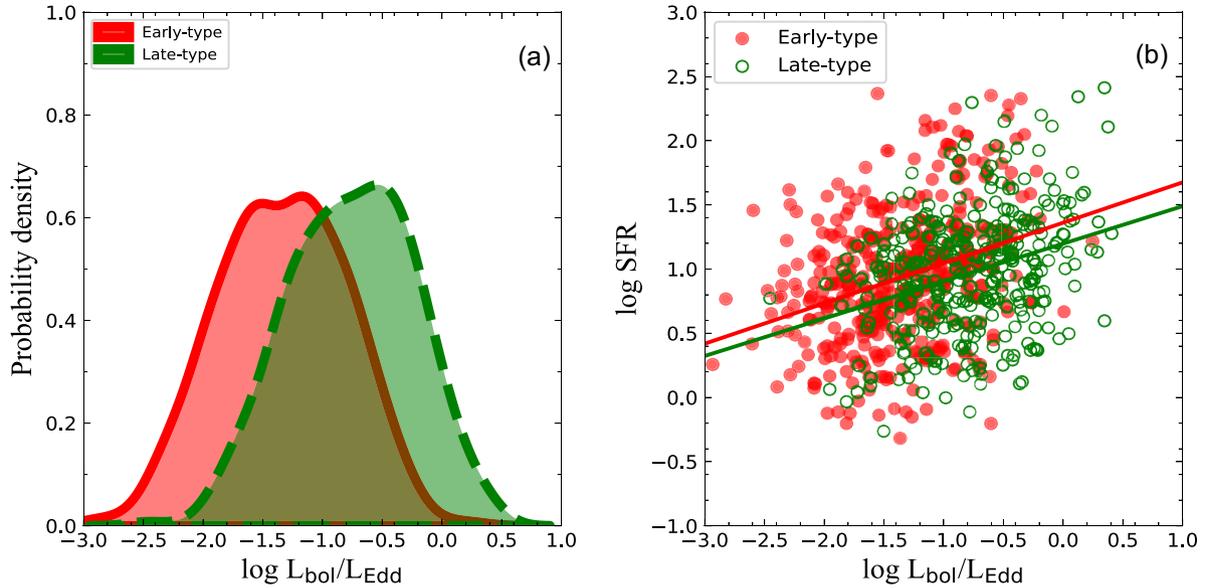

**Figure 5.** (a) The distribution of Eddington ratios for massive star-forming galaxies. (b) The relation between Eddington ratio and SFR for massive star-forming galaxies. The red solid line is the best-fitting line for massive star-forming early-type galaxies. The green solid line is the best-fittng line for massive star-forming late-type galaxies. The red dot is a massive star-forming early-type galaxies. The green dot is a massive star-forming late-type galaxies.

The relation between Eddington ratio and SFR for massive star-forming galaxies is shown in the right panel of Fig. 5. We find a significant correlation between the Eddington ratio and SFR for massive star-forming early-type galaxies ($r = 0.32$ and $p = 2.37 \times 10^{-11}$),

$$\log \text{SFR} = (0.31 \pm 0.05) \log L_{\text{bol}}/L_{\text{Edd}} + (1.36 \pm 0.07). \quad (9)$$

There is also a significant correlation between Eddington ratio and SFR for massive star-forming late-type galaxies ($r = 0.32$ and $p = 1.60 \times 10^{-8}$),

$$\log \text{SFR} = (0.29 \pm 0.05) \log L_{\text{bol}}/L_{\text{Edd}} + (1.19 \pm 0.05). \quad (10)$$

Zhuang & Ho (2020) also discovered that the SFR depends on the Eddington ratio for type I AGN. From equations (9) and (10), we also find that the slope of the relationship between SFR and Eddington ratio in early- and late-type galaxies is consistent within the error range. The accretion rate serves as an indicator of black hole activity. These findings suggest that the influence of black hole activity on star formation is similar in both early- and late-type galaxies.

At the same time, we find that nearly all of the massive star-forming gaaxies in our study have Eddington ratios $\log L_{\text{bol}}/L_{\text{Edd}} > 2.0$. Recent three-dimensional radiation magnetohydrodynamic simulations suggest that the accretion disc with high Eddington ratios can produce high-speed outflows (Jiang, Stone & Davis 2019a; Jiang et al. 2019b). This theoretical framework offers a plausible explanation for the observed correlation between SFR and Eddington ratio. The outflow generated by the accretion disc with high Eddington ratios may provide a source of positive feedback to the interstellar medium of the host galaxies. Compression of cold molecular gas (Silk 2013) or direct star formation in the outflow (Ishibashi & Fabian 2012; Ishibashi, Fabian & Canning 2013; Maiolino et al. 2017; Gallagher et al. 2019) can promote star formation.

## 4 CONCLUSIONS

In this article, we mainly investigate the impact of feedback from AGN on massive star-forming galaxies. We use radio luminosity to calculate the spin of black holes and study the relationship between black hole spin and star formation. Our main conclusions are as follows:

(1) The late-type galaxies tend to have higher average SFRs, specific SFRs, and black hole spin than early-type galaxies. The early-type galaxies tend to have higher average black hole mass and star formation activity parameters than late-type galaxies.

(2) There is a significant correlation between black hole spin and SFR, specific SFR, and star formation activity parameter for massive star-forming early- and late-type galaxies. These results indicate that the spin of supermassive black holes regulates the star formation of massive star-forming early- and late-type galaxies.

(3) In early- and late-type galaxies, the slopes of the relationship between black hole spin and SFR, specific SFR, and star formation activity parameters are consistent within the error range. These results suggest that the mechanism of black hole spin regulating star formation may be similar in early- and late-type galaxies.

**ACKNOWLEDGEMENTS**

YC is grateful for financial support from the National Natural Science Foundation of China (no. 12203028). YC is grateful for funding for the training Program for talents in Xingdian, Yunnan Province (2081450001). QSGU is supported by the National Natural Science Foundation of China (12121003, 12192220, and 12192222). We also acknowledge the science research grants from the China Manned Space Project with no. CMS-CSST-2021-A05. This work is supported by the National Natural Science Foundation of China (11733001, U2031201 and 12433004). XG acknowledge the support of National Nature Science Foundation of China (nos 12303017). This work is also supported by Anhui Provincial Natural Science Foundation project no. 2308085QA33. DRX is supported by the NSFC 12473020, Yunnan Province Youth Top Talent Project (YNWR-QNBJ-2020-116), and the CAS Light of West China Program.

## DATA AVAILABILITY

All the data used here are available upon reasonable request. All data are in Table 1.

## SUPPORTING INFORMATION

Supplementary data are available at *MNRAS* online.

Please note: Oxford University Press is not responsible for the content or functionality of any supporting materials supplied by the authors. Any queries (other than missing material) should be directed to the corresponding author for the article.

This paper has been typeset from a T<sub>E</sub>X/LAT<sub>E</sub>X file prepared by the author.